\documentclass[10pt,aps,prl,twocolumn,showpacs,showkeys]{revtex4-1}

\usepackage{graphicx}
\usepackage{dcolumn}
\usepackage{bm}
\usepackage{color}
\begin{document}

\title{Role of Spin Diffusion in Current-Induced Domain Wall Motion}
\date{\today}
\author{A. Manchon$^1$}
\email{aurelien.manchon@kaust.edu.sa}
\author{W.-S. Kim$^2$}
\author{K.-J. Lee$^{2,3}$}
\affiliation{$^1$Materials Science and Engineering, Division of Physical Science and Engineering, KAUST, Thuwal 23955, Saudi Arabia\\$^2$Department of Materials Science and Engineering, Korea University, Seoul 136-713, Korea\\$^{3}$Electron Physics Group, National Institute of Standards and Technology, Gaithersburg, Maryland 20899-8412, USA}
\begin{abstract}
Current-induced spin torque and magnetization dynamics in the presence of spin diffusion in magnetic textures is studied theoretically. We uncover an additional torque on the form $\sim{\bm\nabla}^2[{\bf M}\times({\bf u}\cdot{\bm \nabla}){\bf M}]$, where ${\bf M}$ is the local magnetization and ${\bf u}$ is the direction of injected current. This torque is inversely proportional to the square of the domain wall width ($\approx\frac{1}{W^2}$) and strongly depends on the domain wall structure. Whereas its influence remains moderate for transverse domain walls, it can significantly increase the transverse velocity of vortex cores. Consequently, the spin diffusion can dramatically enhance the non-adiabaticity of vortex walls.
\end{abstract}
\pacs{72.25.-b,75.60.Ch}
\maketitle
The electrical control of the magnetic state of nanoscale heterostructures \cite{stt} such as magnetic domain walls \cite{dwm} and vortex cores \cite{vortex} is attracting increasing interest as a promising mechanism for innovative memory devices \cite{racetrack}. Identifying the nature of the torque exerted by the injected current on the domain wall itself has constituted a stimulating challenge resulting in the observation of unique dynamical behaviors \cite{dwm,vortex,racetrack,dyn,thomas} and raising seminal questions concerning the transport of itinerant spins in inhomogeneous magnetic textures \cite{zhangli,thiaville,tatara,xiao,bohlens,zz}. The most widely accepted form of the spin transfer torque exerted by a charge current on a magnetic texture ${\bf M}({\bf r},t)$ is \cite{zhangli,thiaville}
\begin{equation}
{\bf T}=b_J({\bf u}\cdot{\bm\nabla}){\bf M}-\beta \frac{b_J}{M_s}{\bf M}\times({\bf u}\cdot{\bm\nabla}){\bf M}
\end{equation}
where $b_J$ is the {\em adiabatic} spin torque, $\beta$ describes the non-adiabaticity of the spin torque and ${\bf u}$ is the direction of current injection. The non-adiabaticity $\beta$ is generated by different mechanisms such as spin relaxation \cite{zhangli,bohlens} and magnetic texture-induced spin mistracking \cite{tatara,xiao,bohlens}. In addition to non-adiabaticity, it has recently been found that the Gilbert damping $\alpha$ can also be affected (enhanced) by the spin texture \cite{zz}. From the viewpoint of domain wall dynamics, the longitudinal (transverse) velocity of transverse domain walls (vortex cores) is controlled by the ratio of $\beta$ to $\alpha$ \cite{thiaville}. Therefore, experimental efforts have been expended in accurately determining $\beta$ and $\alpha$ for a wide range of magnetic materials and domain wall widths \cite{dwm,vortex,racetrack,dyn,thomas}.\par

Recent experiments have shown that this ratio depends on the domain wall structure (Bloch, N\'eel or Vortex wall). Thomas \textit{et al.} \cite{thomas}, Eltschka \textit{et al.} \cite{Eltschka} and Heyne \textit{et al.} \cite{Heyne} have found that vortex cores exhibit a much larger non-adiabaticity ($\beta\approx 8\alpha$ to 10$\alpha$) compared to transverse domain walls ($\beta\approx \alpha$). The authors attribute these large non-adiabaticities to the narrow character of domain wall width in vortex cores ($\approx$10 nm). As a matter of fact, Tatara \textit{et al.} \cite{tatara} and Xiao \textit{et al.} \cite{xiao} demonstrated that in sharp domain walls the itinerant spin cannot adiabatically follow the local spin texture, resulting in an enhancement of the non-adiabaticity. On the other hand, Burrowes \textit{et al}. \cite{Burrowes} have tested a very sharp transverse wall of about 1 nm using FePt nanowires and found that such a narrow domain wall does not cause a significant increase in the non-adiabaticity, $\beta\approx\alpha$. Theoretical investigations by Bohlens and Pfannkuche \cite{bohlens} recently showed that the non-adiabatic torque has damped oscillatory behavior when increasing the domain wall width, which may account for the non-adiabaticity enhancement for sharp domain wall widths. However, this model is applied to Bloch walls only and does not readily explain the observed differences between transverse walls and vortex cores.\par

The experimental observations \cite{thomas,Eltschka,Heyne,Burrowes} indicate that the nature of the non-adiabatic spin torque exerted on sharp magnetization patterns is related to their dimensionality. Indeed, whereas a magnetic transverse wall varies along one direction only ($\partial_x{\bf M}\neq0,\;\partial_y{\bf M}=0$), the magnetization of a vortex core varies along two directions ($\partial_x{\bf M}\neq0,\;\partial_y{\bf M}\neq0$). Therefore, one approach to explain the different non-adiabaticities of sharp transverse domain walls and vortex cores is to consider a mechanism that couples both $x$ and $y$ directions. We have recently shown that a transverse spin current caused by anomalous Hall effect increases the transverse velocity of isolated vortex cores while leaving the transverse domain walls essentially unchanged \cite{apl}. However, its contribution is of the order of the damping constant $\alpha$, which is insufficient to explain the results observed in Refs.~\cite{thomas,Eltschka,Heyne}.\par

In this letter, we demonstrate that spin diffusion gives rise to an additional spin torque that contributes to the current-driven velocity in the case of both Bloch walls and isolated vortex cores. However, whereas this additional component affects the velocities of Bloch walls only moderately, it can dramatically increase the velocities of vortex cores. The itinerant electrons evolving in magnetic textures are described by the one-electron Hamiltonian
\begin{equation}
{\hat H}=\frac{{\hat{\bf p}}^2}{2m}+\frac{J_{\rm ex}}{M_s}{\hat{\bm \sigma}}\cdot{\hat{\bf M}}({\bf r},t),
\end{equation}
where $\hat{\bf p}$ is the momentum, $\hat{\bm \sigma}$ is the Pauli matrix, ${\hat{\bf M}}({\bf r},t)$ is the time- and spatial-dependent magnetization ($|{\bf M}|=M_s$) and $J_{\rm ex}$ is the exchange coupling energy. Using Ehrenfest relation $\partial_t{\bf m}=\partial_t\langle{\hat{\bm \sigma}}\rangle=i/\hbar\langle[{\hat{\bm \sigma}},{\hat H}]\rangle$, we obtain the spin continuity equation for the spin density ${\bf m}$
\begin{eqnarray}\label{eq:spincont}
&&\partial_t {\bf m}=-{\bm \nabla}\cdot{\cal J}-\frac{1}{\tau_{\rm ex}M_s}\delta{\bf m}\times{\bf M}-\frac{\delta{\bf m}}{\tau_{\rm sf}},
\end{eqnarray}
where ${\cal J}=\langle{\bf v}\otimes{\bm \sigma}\rangle$ is the spin current tensor, $\tau_{\rm ex}=\hbar/2J_{ex}$  is the spin precession time ($\approx 10^{-15}$ s to $10^{-14}$ s), $\tau_{\rm sf}$ is the phenomenological spin-flip relaxation time ($\approx 10^{-13}$ s to $10^{-12}$ s). We assume ${\bf m}=\frac{n_s}{M_s}{\bf M}+\delta {\bf m}$, $n_s$ ($\delta{\bf m}$) being the (non-)equilibrium spin density ($n_s\approx 10^{-2}M_s$).\par

Notice that Eq. (\ref{eq:spincont}) should in principle include spin dephasing contribution, arising from the destructive interference of non-equilibrium spins with different wave vector direction. Microscopic investigations using realistic Fermi surfaces have shown that this effect destroys the transverse component of itinerant spins within a few monolayers in strong ferromagnets \cite{spindephasing}. Disregarding this mechanism renders the algebra more tractable without qualitatively modifying the conclusion of the present Letter.

As mentioned above, a promising candidate to explain the experimental observations is a mechanism that would link both $x$-$y$ directions. In diffusive systems such as ferromagnetic metallic nanowires, the local momentum scattering tends to counteract the influence of the external electric field resulting in an additional diffusion term in both charge and spin currents \cite{diff} that couples $x$ and $y$ directions. Moreover, in order to track possible influence of sharp magnetic textures on the damping constant $\alpha$, we also incorporate the correction proposed by Zhang and Zhang \cite{zz}. These authors have shown that in a time- and spatial-dependent magnetic texture, local spin pumping (or spin-motive force - SMF) results in a spatial-dependent tensor form of the damping \cite{zz}.
In the drift-diffusion approximation, in the presence of a local spin motive force (see Ref. \cite{zz} for details), the $i$-th component of the spin current reads
\begin{eqnarray}\label{eq:spincurrent}
&&{\cal J}_i=-b_Ju_i{\bf M}-D_0\partial_i\delta{\bf m}+\frac{\eta}{M_s}\partial_t {\bf M}\times\partial_i{\bf M}.
\end{eqnarray}
Here $b_J=\mu_BPG_0E/eM_s$ and $\eta=g\mu_B\hbar G_0/4e^2M_s$, $P$ is the spin polarization, $\mu_B$ is the Bohr magnetron, $G_0$ the electrical conductivity, $D_0$ is the diffusion coefficient and $u_i$ is the $i$-th direction of the injected current. In Eq. (\ref{eq:spincurrent}), the first term is the adiabatic spin current driven by the current density ${\bf j}=G_0E{\bf u}$, the second term arises from carrier diffusion and the last term arises from the SMF \cite{zz}. Notice that spin diffusion itself reduces the magnitude of SMF-damping depending on the spin diffusion length and the domain wall width \cite{zz2,kim}. However, the analytical treatment of this effect is complex, so in the present work we assume a moderate value of the SMF parameter $\eta$ to qualitatively account for this reduction.\par

By injecting Eq. (\ref{eq:spincurrent}) into Eq. (\ref{eq:spincont}), we obtain the diffusion equation for the non-equilibrium spin density $\delta{\bf m}$
\begin{eqnarray}\label{eq:spin1}
&&\partial_t \delta{\bf m}-D_0{\bm\nabla}^2\delta{\bf m}+\frac{1}{\tau_{\rm ex}M_s}\delta{\bf m}\times{\bf M}+\frac{\delta{\bf m}}{\tau_{\rm sf}}=\\
&&b_J({\bf u}\cdot{\bm \nabla}){\bf M}+\frac{\eta}{M_s}{\bf M}\times({\cal A}\cdot\partial_t{\bf M})-\frac{n_s}{M_s}\partial_t{\bf M},\nonumber
\end{eqnarray}
where ${\cal A}$ is the tensor defined by ${\cal A}_{\rm\alpha\beta}=\frac{1}{M_s^4}\sum_i({\bf M}\times\partial_i{\bf M})_\alpha({\bf M}\times\partial_i{\bf M})_\beta$ \cite{zz}. The right hand side of Eq. (\ref{eq:spin1}) acts as a source of itinerant spin dynamics inside the domain wall. Note that the spin diffusion ($\propto{\bm\nabla}^2\delta{\bf m}$) was neglected in Ref. \cite{zhangli}. In Eq. (\ref{eq:spin1}), by taking $\partial_t\sim1/\tau_M$ and ${\bm\nabla}\sim1/W$ ($\tau_M$ and $W$ are respectively the domain wall dynamics timescale and width), we have $1/\tau_{\rm ex},1/\tau_{\rm sf},b_J/W>>1/\tau_M,\eta/(\tau_M W^2), D_0/W^2$. Therefore, the left-hand side of Eq. (\ref{eq:spin1}) is dominated by the torque and relaxation terms ($\propto1/\tau_{\rm ex},1/\tau_{\rm sf}$) and the lowest order itinerant spin density reduces to \cite{zhangli,thiaville}
\begin{eqnarray}\label{eq:ad}
\delta{\bf m}&\approx &\frac{\tau_{\rm ex}}{1+\beta^2}\frac{b_J}{M_s}(1-\frac{\beta}{M_s}{\bf M}\times){\bf M}\times({\bf u}\cdot{\bm \nabla}){\bf M}\nonumber\\
&&+\tau_{\rm ex}{\cal O}\left(\frac{1}{\tau_M},\frac{D_0}{W^2},\frac{\eta}{\tau_M W^2}\right),
\end{eqnarray}
where $\beta=\tau_{\rm ex}/\tau_{\rm sf}$. By injecting Eq. (\ref{eq:ad}) into Eq. (\ref{eq:spin1}), we get
\begin{eqnarray}\label{eq:spin2}
&&\frac{1}{\tau_{\rm ex}M_s}\delta{\bf m}\times{\bf M}+\frac{\delta{\bf m}}{\tau_{\rm sf}}=b_J({\bf u}\cdot{\bm \nabla}){\bf M}-\frac{n_s}{M_s}\partial_t{\bf M}\\
&&+\frac{\eta}{M_s}{\bf M}\times({\cal A}\cdot\partial_t{\bf M})+\lambda_{\rm ex}^2\frac{b_J}{M_s}{\bm\nabla}^2\left[{\bf M}\times({\bf u}\cdot{\bm \nabla}){\bf M}\right]\nonumber
\end{eqnarray}

Here $\lambda_{\rm ex}^2=D_0\tau_{\rm ex}/(1+\beta^2)$ is the transverse spin diffusion length already mentioned in metallic spin-valves \cite{zlf}. The term $\partial_t \delta{\bf m}$ ($\propto\tau_{\rm ex}/\tau_{\rm M}$) as well as the highest order terms ($\propto\beta^2,\beta\eta,\lambda_{\rm ex}^2\beta, \lambda_{\rm ex}^2\eta<<\beta$) have been disregarded since we retain only terms comparable to $\beta$. Equation (\ref{eq:spin2}) can be easily manipulated to get the first order corrections to the spin torque exerted on the magnetic texture \cite{zhangli}, ${\bf T}=\frac{1}{\tau_{\rm ex}M_s}\delta{\bf m}\times{\bf M}={\bf T}_{\rm ren}+{\bf T}_{\rm st}+{\bf T}_{\rm d}$, where
\begin{eqnarray}\label{eq:spin3}
{\bf T}_{\rm ren}&=&-\frac{n_s}{M_s}\partial_t{\bf M}+\beta\frac{n_s}{M_s^2}{\bf M}\times\partial_t{\bf M},\\
{\bf T}_{\rm st}&=&b_J({\bf u}\cdot{\bm \nabla}){\bf M}-\beta \frac{b_J}{M_s}{\bf M}\times({\bf u}\cdot{\bm \nabla}){\bf M}\nonumber\\
&&+\lambda_{\rm ex}^2\frac{b_j}{M_s}{\bm\nabla}^2[{\bf M}\times({\bf u}\cdot{\bm \nabla}){\bf M}],\\
\label{eq:spin4}
{\bf T}_{\rm d}&=&\frac{\eta}{M_s}{\bf M}\times({\cal A}\cdot\partial_t{\bf M}).
\end{eqnarray}
Therefore, the torque can be decomposed into three contributions: renormalization torque ($\propto\partial_t{\bf M}$), spin torque ($\propto({\bf u}\cdot{\bm \nabla}){\bf M}$) and damping torque ($\propto {\cal A}\cdot\partial_t{\bf M}$). In the remaining of this work, we neglect the renormalization torque ($n_s<<M_s$).\par

In the following we study the combined influence of the spin diffusion and SMF-induced damping on two typical magnetization patterns: a Bloch wall and a free vortex core. The magnetization dynamics is governed by the Landau-Lifshitz-Gilbert (LLG) equation
\begin{eqnarray}\label{eq:LLG}
\partial_t{\bf M}&=&-\gamma{\bf M}\times{\bf H}_{\rm eff}+\frac{\alpha}{M_s}{\bf M}\times\partial_t{\bf M}+{\bf T}_{\rm d}+{\bf T}_{\rm st},\\
{\bf H}_{\rm eff}&=&H_K\frac{M_x}{M_s}{\bf e}_x+2\frac{A}{M_s^2}{\bm\nabla }^2{\bf M}-4\pi \frac{M_z}{M_s}{\bf e}_z+{\bf H},
\end{eqnarray}
where $H_K$ is the uniaxial anisotropy field, A is the exchange constant, and ${\bf H}$ is the external applied field. In a perpendicularly magnetized nanowire, a {\em Bloch wall} is described by $\theta(x)=2\tan^{-1}e^{x/W},\phi=\phi(t)$. The coupled dynamics gives
\begin{eqnarray}\label{eq:coupled}
&&\partial_\tau X=\gamma W\left[\frac{H_K}{2}\sin2\phi+(\alpha+\frac{2\eta}{3W^2})H_z\right]\nonumber\\
&&-b_J\left[1+\beta(\alpha+\frac{2\eta}{3W^2})+\frac{\lambda_{\rm ex}^2}{W^2}(\frac{\alpha}{3}+\frac{2\eta}{5W^2})\right],\\
&&\partial_\tau\phi=\gamma \left[H_z-\alpha\gamma \frac{H_K}{2}\sin2\phi\right]+\frac{b_J}{W}\left[\alpha-\beta-\frac{1}{3}\frac{\lambda_{\rm ex}^2}{W^2}\right],
\end{eqnarray}
where $\tau=t\left(1+\alpha(\alpha+\frac{2\eta}{3W^2})\right)$, and $X$ is the position of the domain wall center. At the lowest order, we obtain the velocity below and above the Walker breakdown
\begin{eqnarray}\label{eq:v_bb}
&&{v}_x^{<WB}\approx\frac{W \gamma H_z-b_J(\beta+\frac{\lambda_{\rm ex}^2}{2W^2})}{\alpha(1+\alpha^2+\eta')},\\
&&{v}_x^{>WB}\approx\frac{\gamma W(\alpha+\eta')H_z-b_J}{1+\alpha^2+\eta'},
\end{eqnarray}
where $\eta'=2\alpha\eta/3\Delta^2$ is the reduced SMF parameter. The spin diffusion only significantly affects the transverse wall velocity below the Walker breakdown by enhancing the effective non-adiabaticity.\par
Alternatively, in a two-dimensional magnetic stripe, an {\em isolated vortex core} is described by $\theta(x,y)=2\tan^{-1}r/r_0$ for $r=\sqrt{x^2+y^2}\leq r_0$, $\theta=\pi/2$ for $r_0\leq r\leq R$, and $\phi=\tan^{-1}y/x+\pi/2$, where $r_0$ ($R$) is the inner (outer) radius of the vortex core. We use Thiele's description of rigid domain wall motion \cite{thiele}, where $\partial_t{\bf M}=-({\bf v}\cdot{\bm\nabla}){\bf M}$. Thiele has shown that Eq. (\ref{eq:LLG}) can be expressed in the form of the sum of forces exerted on the wall \cite{thiele}. By multiplying Eq. (\ref{eq:LLG}) on the left by $\frac{{\bf M}}{M_s}\times$ and projecting the obtained equation on $-\partial_i{\bf M}/\gamma M_s$, one obtains
\begin{eqnarray}\label{eq:forces}
&&\int_\Omega \left[-{\bf H}_{\rm eff}\cdot\partial_i{\bf M}+{\bf G}\times({\bf v}+b_J{\bf u})+{\cal D}\cdot(\alpha{\bf v}+\beta
b_J{\bf u})\right.\nonumber\\
&&\left.-\frac{b_J}{\gamma M_s^3}\lambda_{\rm ex}^2[{\bf M}\times{\bm\nabla}^2({\bf M}\times({\bf u}\cdot{\bm\nabla}){\bf M})]\cdot\partial_i{\bf M}\right.\nonumber\\
&&-\left.\frac{\eta}{\gamma M_s}[{\cal A}\cdot({\bf v}\cdot{\bm\nabla}){\bf M}]\cdot\partial_i{\bf M}\right] d\Omega=0\\
&&{\bf G}=-\frac{M_s}{\gamma}\sin\theta({\bm\nabla}\theta\times{\bm\nabla}\phi)\\
&&{\cal D}_{\rm ij}=-\frac{M_s}{\gamma}(\partial_i\phi\partial_j\phi\sin^2\theta+\partial_i\theta\partial_j\theta)
\end{eqnarray}
The integral $\int_\Omega d\Omega$ runs over the volume $\Omega$ of the magnetic stripe. After some algebra, the longitudinal and transverse velocities of the vortex core read
\begin{eqnarray}
v_x&=&-\frac{4+\alpha_{\rm eff}\beta_{\rm eff}}{4+\alpha_{\rm eff}^2}b_j, v_y=-2\frac{\alpha_{\rm eff}-\beta_{\rm eff}}{4+\alpha_{\rm eff}^2}b_j\label{eq:vy1}
\end{eqnarray}
where we define the effective damping as $\alpha_{\rm eff}={\cal C}\alpha+\frac{14\eta}{3r_0^2}$, the effective non-adiabaticity as $\beta_{\rm eff}={\cal C}\beta+\frac{28\lambda_{\rm ex}^2}{3r_0^2}$, 
 and ${\cal C}=2+\ln\frac{R}{r_0}$.\par

Fig. 1 displays the Bloch wall velocities as a function of the domain wall width $W$. The velocities are normalized to the case without spin diffusion or SMF-damping ($\lambda_{\rm ex}=\eta=0$). Below Walker breakdown [Fig. 1(a)], the velocity is moderately affected by the domain wall width. For example, when $W=5$ nm, the normalized velocity increases by a factor of 2 for $\lambda_{\rm ex}=0.8$ nm (equivalent to NiFe with $J_{\rm ex}\approx0.5$ eV, see also Ref. \cite{urazh}). The SMF-damping is almost absent in this regime (see dots in Fig. 1). Above Walker breakdown [Fig. 1(b)], the velocity is simply not affected by the domain wall width (less than 0.01 \%). \par



\begin{figure}
	\centering
		\includegraphics[width=7cm]{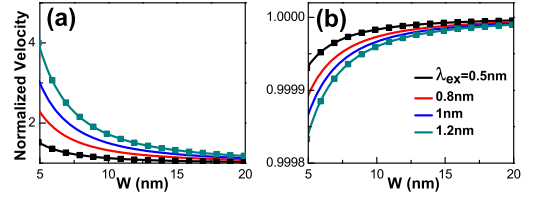}
	\label{fig:Fig1a}\caption{(Color online) Normalized velocity of the Bloch wall, below ${v}_x^{<WB}$ (a) and above Walker breakdown ${v}_x^{>WB}$ (b). The parameters are $\alpha=0.005$, $\beta=0.01$ and $\eta=0.2$ nm$^2$. The dots represent the normalized velocities in the presence of SMF-damping.}
\end{figure}

The case of vortex cores is very different. Fig. 2(a) and (b) display the longitudinal and transverse velocities of an isolated vortex core as a function of the core radius $r_0$. While the longitudinal velocity is not significantly affected, the transverse velocity is dramatically enhanced by the presence of diffusion. For $r_0=5$ nm, the transverse velocity can be increased by a factor of 10 for a transverse spin diffusion length $\lambda_{\rm ex}=0.8$ nm. On the other hand, accounting for a moderate SMF-damping, $\eta=0.2$ nm$^2$, the normalized velocity can be slightly reduced. This observation has very important implications in the evaluation of the non-adiabatic torque in vortex walls. It indicates that the traditional way to extract $\beta$ must be reconsidered and that a more complete analytical treatment of the velocities [i.e. Eq. (\ref{eq:vy1})] needs to be performed. One way to extract the non-adiabaticity parameter is to estimate the polar angle $\tan^{-1}v_y/v_x$ acquired by the isolated vortex core after current injection \cite{Heyne}. Fig. 3(a) shows that this angle can be also dramatically enhanced at small core sizes due to the transverse spin diffusion.\par
\begin{figure}
	\centering
		\includegraphics[width=7cm]{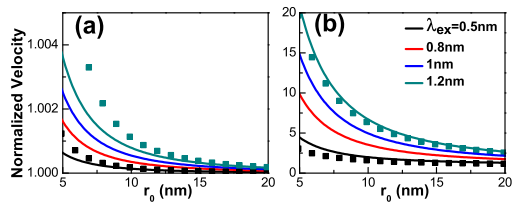}
	\label{fig:Fig1b}\caption{(Color online) Normalized longitudinal $v_x$ (a) and transverse $v_y$ (b) velocity of the isolated vortex core. The parameters are $R=150$ nm, $\alpha=0.005$, $\beta=0.01$ and $\eta=0.2$ nm$^2$. The dots represent the normalized velocities in the presence of SMF-damping.}
\end{figure}

Up until now, the mechanism that was expected to mostly contribute to the non-adiabatic torque for sharp domain walls ($W<10$ nm) has been the ballistic spin mistracking proposed in Refs. \cite{tatara,xiao}. This effect is non-local and increases with decreasing the exchange and domain wall width. To assess the importance of the spin diffusion mechanism compared to the ballistic spin mistracking mechanism, we numerically calculated the non-adiabaticity caused by the ballistic spin mistracking based on Ref. \cite{xiao}. The non-adiabaticity parameter evaluated numerically is defined as the ratio between the non-adiabatic torque and the adiabatic torque at the center of the wall. As displayed in Fig. 3(b), the estimated contribution of the ballistic spin mistracking to the non-adiabaticity remains very limited. Since most of the ferromagnetic materials used in experiments are strong ferromagnets ($\lambda_{\rm ex}<0.5$ nm), this torque only has a sizable influence for extremely sharp domain walls ($W\leq1$ nm). Therefore, for moderately sharp domain walls such as vortex cores, the transverse spin diffusion would give the most important contribution to the non-adiabaticity.\par
Finally, let us discuss about the relevance of our results to previous experimental data. Our calculations [see Fig. 2(b) and 3(a)] are consistent with the large non-adiabaticities measured in Refs. \cite{thomas,Eltschka,Heyne} for vortex cores in NiFe. In these structures the transverse spin diffusion length is about $\lambda_{\rm ex}\approx0.8$ nm \cite{spindephasing,urazh} and the radius $r_0\approx5$ nm, which yields effective velocities about 10 times larger than in the absence of spin diffusion, as mentioned above. On the other hand, the very narrow domain walls investigated by Burrowes \textit{et al.} \cite{Burrowes} are obtained for FePt, which is a strong ferromagnet with a very short transverse spin diffusion length, $\lambda_{\rm ex}\leq0.5$ nm. In this case, as shown in Fig. 1(a) and Fig. 3(b), both spin diffusion and ballistic mistracking contributions are quenched. For strong ferromagnets, the itinerant electron spin is very rapidly aligned on the local magnetization and all effects related to spin misalignment vanish. As a consequence, in such materials, non-adiabaticity might be dominated by contributions from the longitudinal spin relaxation \cite{zhangli}.\par
\begin{figure}
	\centering
		\includegraphics[width=6cm]{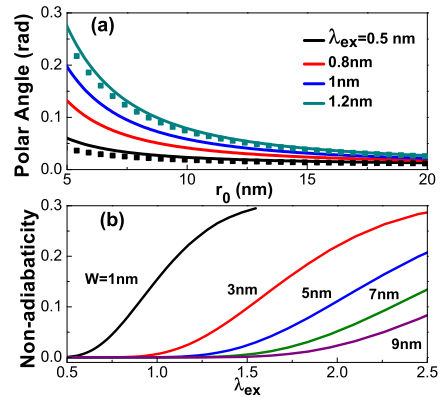}
	\label{fig:Fig2}\caption{(Color online) (a) Polar angle of the vortex core as function of the core radius. The dots represent the normalized angle in the presence of SMF-damping. Same parameters as in Fig. 2; (b) Non-adiabaticity as a function of the transverse spin diffusion length $\lambda_{\rm ex}$ for decreasing domain wall widths, calculated based on Ref. \cite{xiao}.}
\end{figure}
The authors thank M. D. Stiles and X. Waintal for inspiring discussions. K.J.L. acknowledges financial support from NRF/MEST (Grant No. 2010-0023798).

\end{document}